\documentclass{article}
\usepackage[utf8]{inputenc}
\usepackage[affil-it]{authblk}
\usepackage{graphicx}
\usepackage{subfigure}
\usepackage{color}
\usepackage[margin=2cm]{geometry}

\graphicspath{{figs/}}

\title{Fungal photosensors}
\author[1,*]{Alexander E. Beasley\footnote{Corresponding author: alex.beasley@uwe.ac.uk}}
\author[1]{Anna L. Powell}
\author[1]{Andrew Adamatzky}

\affil[1]{Unconventional Computing Laboratory, UWE, Bristol, UK}

\date{\today}

\begin{document}

\maketitle

\begin{abstract}
\noindent
The rapidly developing research field of organic analogue sensors aims to  replace traditional semiconductors with naturally occurring materials. Photosensors, or photodetectors, change their electrical properties in response to the light levels they are exposed to. Organic photosensors can be functionalised to respond to specific wavelengths, from ultra-violet to red light.  Performing cyclic voltammetry on fungal mycelium and fruiting bodies under different lighting conditions shows no appreciable response to changes in lighting condition. However, functionalising the specimen using PEDOT:PSS yields in a photosensor that produces large, instantaneous current spikes when the light conditions change. Future works would look at interfacing this organic photosensor with an appropriate digital back-end for interpreting and processing the response.    


\end{abstract}

\section{Introduction}
\label{sec:introduction}

The world's drive for sustainability dictates a need to develop `green architectures', where buildings are made from natural, recyclable materials and incorporate living matter in their constructions~\cite{mcdonough2002big,kibert2016sustainable}. Recent examples include algae facades~\cite{kim2013beyond,martokusumo2017algae}, buildings incorporating bio-reactors~\cite{wilkinson2016feasibility} and buildings made from pre-fabricated blocks of substrates colonised by fungi~\cite{ross2016your,appels2019fabrication,dahmen2016soft}. Not long ago we proposed growing monolith constructions from live fungal materials, where living mycelium coexist with dry mycelium functionalised with nanoparticles and polymers~\cite{adamatzky2019fungal}. In such monolith constructions, fungi could act as optical, tactile and chemical sensors, fuse and process information and make decisions~\cite{adamatzky2018towards}. 

Functionalisation of living substrates aimed at increasing their sensitivity or conductivity, or imbuing them with novel properties, has been achieved before. Examples include: tuning electrical properties of plants and slime mould with nanoparticles~\cite{gizzie2016hybridising,gizzie2016living}, increasing photosynthetic properties of plants with nanoparticles~\cite{giraldo2014plant,faizan2018zinc} and hybridizing slime mould with conductive polymers~\cite{berzina2015hybrid,dimonte2016physarum}. An important advantage of functionalising living substrates with particles, and/or polymers, is that the substrate will remain functional, as an inanimate electronic device, even when no longer living. Here, in an attempt to advance our ideas of living fungal architectures, sensors and computers, we explore the photosensitive properties of the grey oyster mushroom with aim of designing and prototyping novel organic electronic devices.

In the field of organic electronics, we can use organic material to replace sensors, such as photosensors~\cite{book:analogorganicelectronics, Ocaya2017OrganicPhotosensors, Dickey1999, Hamilton2004polymerphotosensors, Manna2015, Zhaomiyi2019photodetectors}. Organic photosensors are a potentially disruptive technology that lend themselves well to being tuned for specific band absorption~\cite{Natali2016Photodetectors}. Additionally, organic electronics have the potential to self heal as the substrate is a living material~\cite{Cicoira2018Healable}.  Functionalising organic compounds with polymers can create hybrid components that utilise specific properties from the various individual compounds. For example, increasing the capacitance of a multi-layer capacitor using photosensitivity~\cite{Lee2016PhotoAmplification}. Fungi have the capacity to readily transport polymers which can be used to functionalise the specimen. Given these properties, fungi may be a useful substrate for the development of organic photosensors and a digital back-end can be used to interpret and process the signal.

In recent years, there has been significant developments in the field of organic analogue electronics, particularly in the development of organic thin film transistors~\cite{Tokito2018TFT, Endoh2007OLIT, Tang2018OTFT}, and organic LEDs~\cite{Mizukami2018OLED, Sano2019OLED}. It is only natural, therefore, that the progression of this field continues with the development of other key electronic sensing components, such as the photosensor. 

Photosensors change their impedance as a function of the lighting conditions they are exposed to. The change in impedance can then be processed by other circuit elements as a form of input. It is known that fungi are typically sensitive to changes in lighting conditions.  If exposing fungi to sudden and dramatic changes in lighting results in a rapid change in the electrical properties of the fungi then this can change can be interpreted by other electronic devices. The result is a hybrid electronics system that uses fungi as a photosensor.

The paper will explore the use of the fruiting bodies of the grey oyster fungi as photosensors and present the implications of such an application when used in organic electronic systems. The rest of this paper is structured as follows. Section~\ref{sec:experimentation} presents the experimental set up and Sect.~\ref{sec:results} presents the results with discussion. Finally conclusions are drawn in Sect.~\ref{sec:conclusions}.

\section{Experimental method}
\label{sec:experimentation}

\begin{figure}
    \centering
    \includegraphics[width=0.7\textwidth]{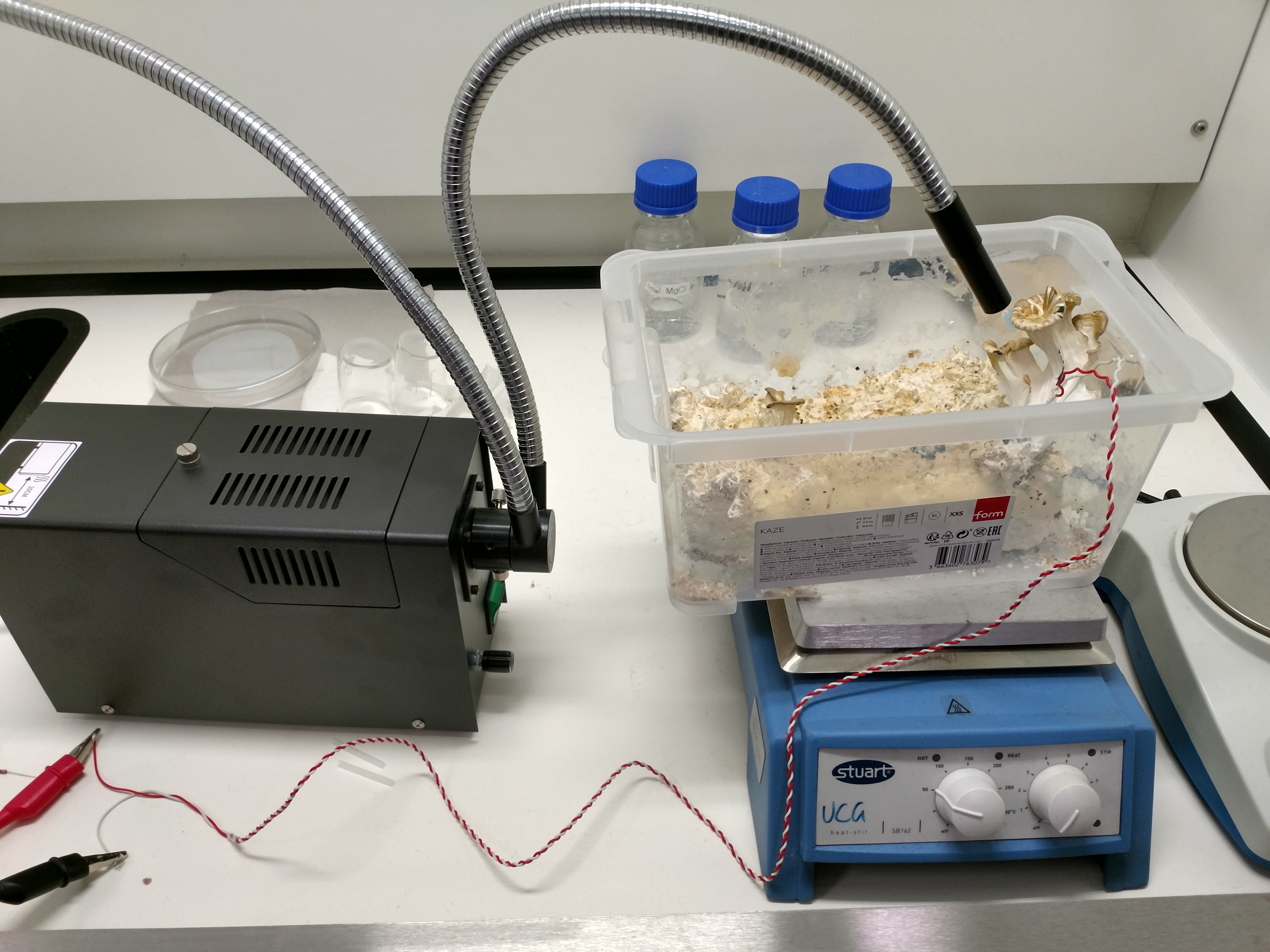}
    \caption{Experimental setup}
    \label{fig:setup}
\end{figure}

Samples of mycelium and fruiting bodies of the grey oyster fungi \emph{Pleurotus ostreatus} (Ann Miller's Speciality Mushrooms Ltd, UK), cultivated on damp wood shavings, are explored for their photosensitive properties. Iridium-coated stainless steel sub-dermal needles with twisted cables (Spes Medica SRL, Italy)  were inserted approximately 10~mm apart in samples (Fig.\ref{fig:setup}). An LCR meter (BK Precision, \emph{model number}) was used to constantly record the DC resistance of the sample. The sample was periodically covered or exposed to intense light (c.~1500\,Lux, with a cold light source PL2000, Photonic Optics, USA).  Exposure to the high intensity light was in the order of tens of seconds. Any changes in resistance were recorded. The container with PEDOT:PSS functionalised fungi was kept in a fume hood for the duration of the experiment.\par

The I-V characteristics of the fungal substrate were measured using a Keithely source measure unit (SMU) 2450 (Keithley Instruments, USA). High intensity light was turned on and off during the course of the cyclic voltammetry to observe instantaneous changes in conducted current. Similar tests were performed with a pure resistive load to ensure no spiking effects were caused by the operation of high intensity equipment. Control conditions are given by performing the same I-V characterisation with constant lighting conditions.\par

In the experiments on functionalised fruit bodies, 10\,ml of PEDOT:PSS (poly(3,4-ethylenedioxythiophene) polystyrene sulfonate) (Sigma Aldrich, UK) solution was injected in the stalks the fruit bodies. The fungi were rested for 24~hr to allow for translocation of them material towards other parts of the stem and caps of the bodies.

\section{Results}
\label{sec:results}

\begin{figure}[!tpb]
    \centering
    \subfigure[]{\includegraphics[width=0.48\textwidth]{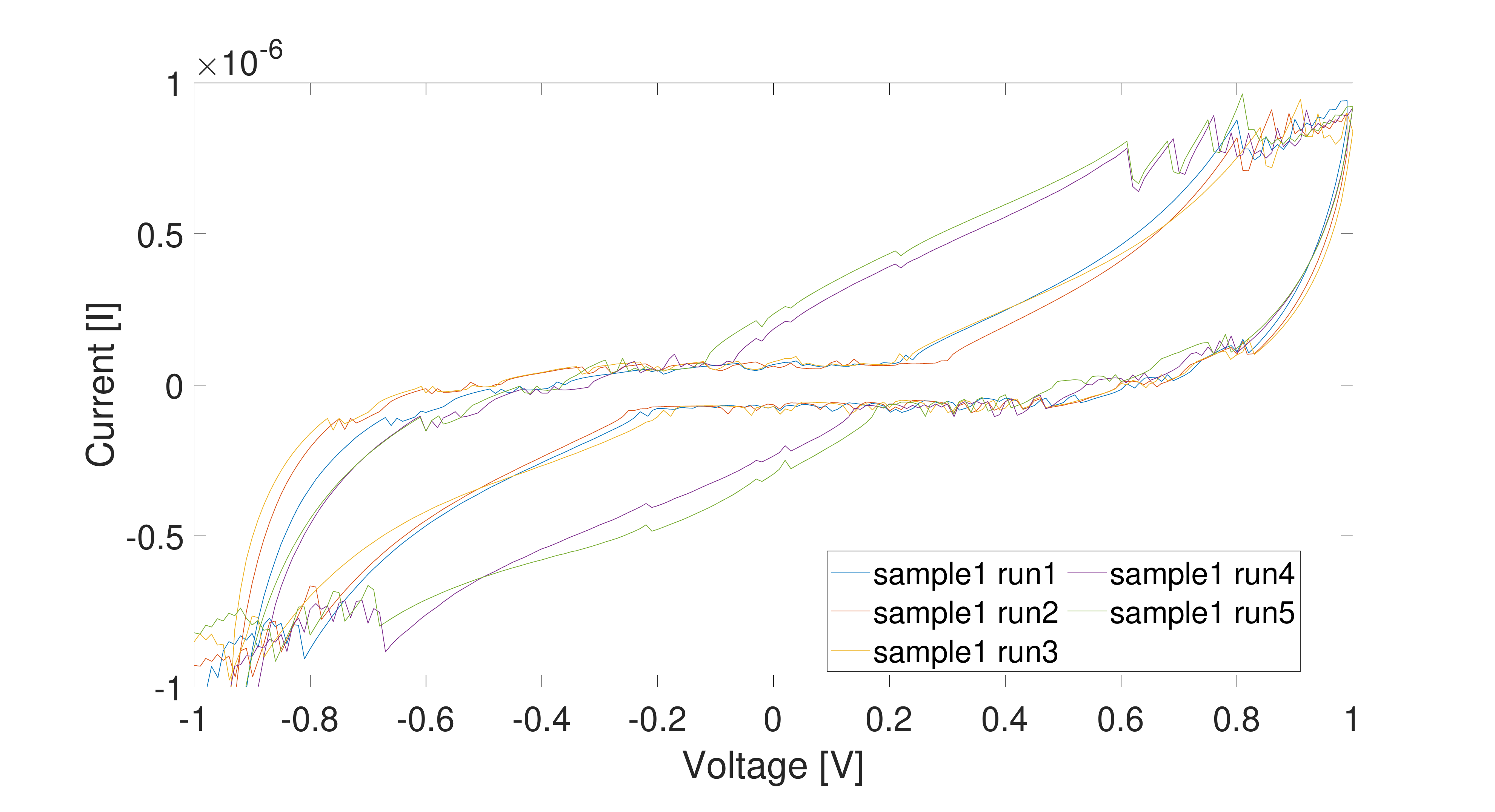}
    \label{fig:sample_1_mycelium}}
    \subfigure[]{\includegraphics[width=0.48\textwidth]{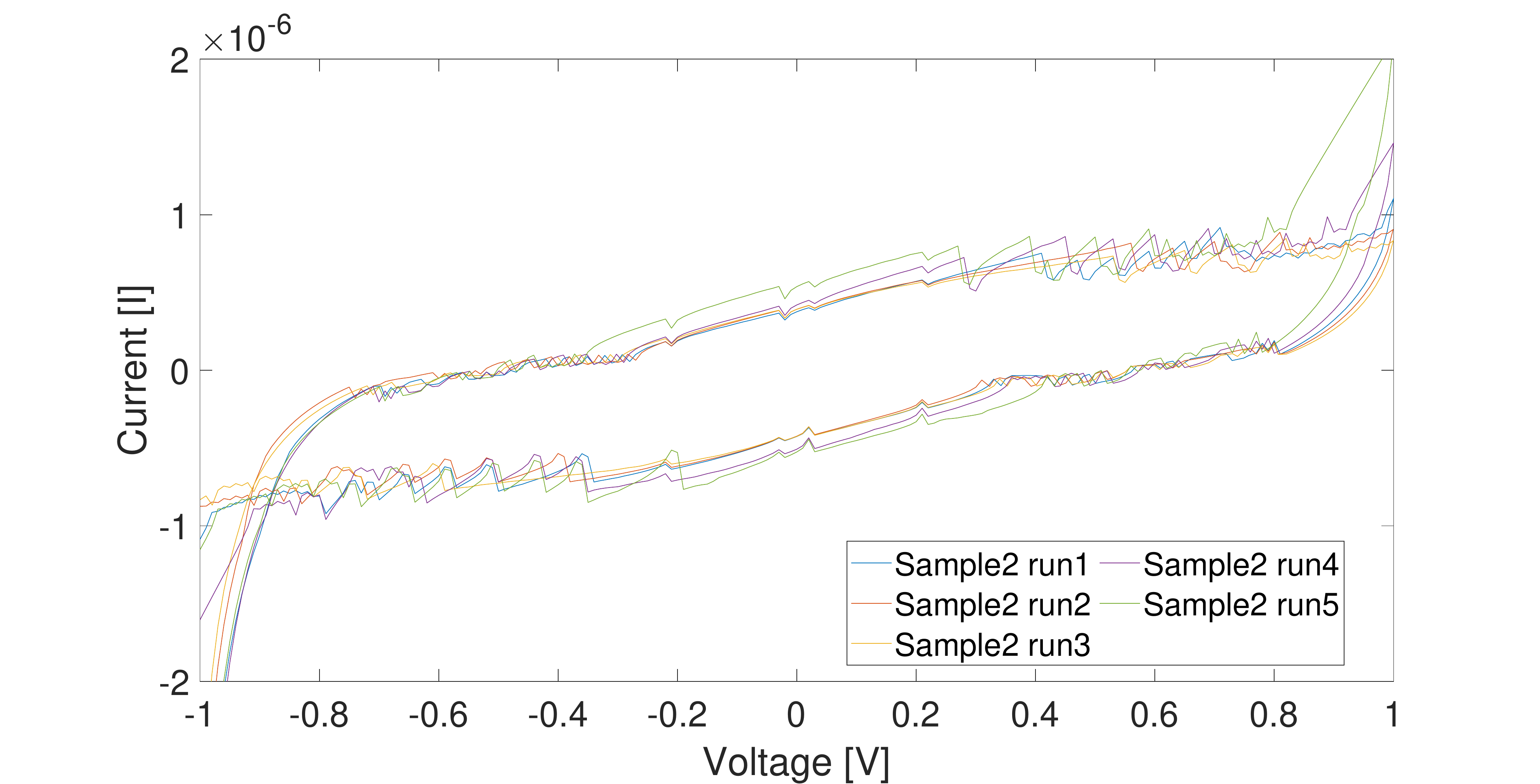}
    \label{fig:sample_2_mycelium}}
    \caption{Cyclic voltammetry of mycelium samples (a) sample 1, (b) sample 2. Run 1 shows the response when the sample is in the dark. Run 2 shows the response of the sample under continuous intense illumination. Runs 3--5 shows the response when the sample is subject to sudden changes in light conditions.}
    \label{fig:pure_mycelium}
\end{figure}

\begin{figure}[!tpb]
    \centering
    \includegraphics[width=0.48\textwidth]{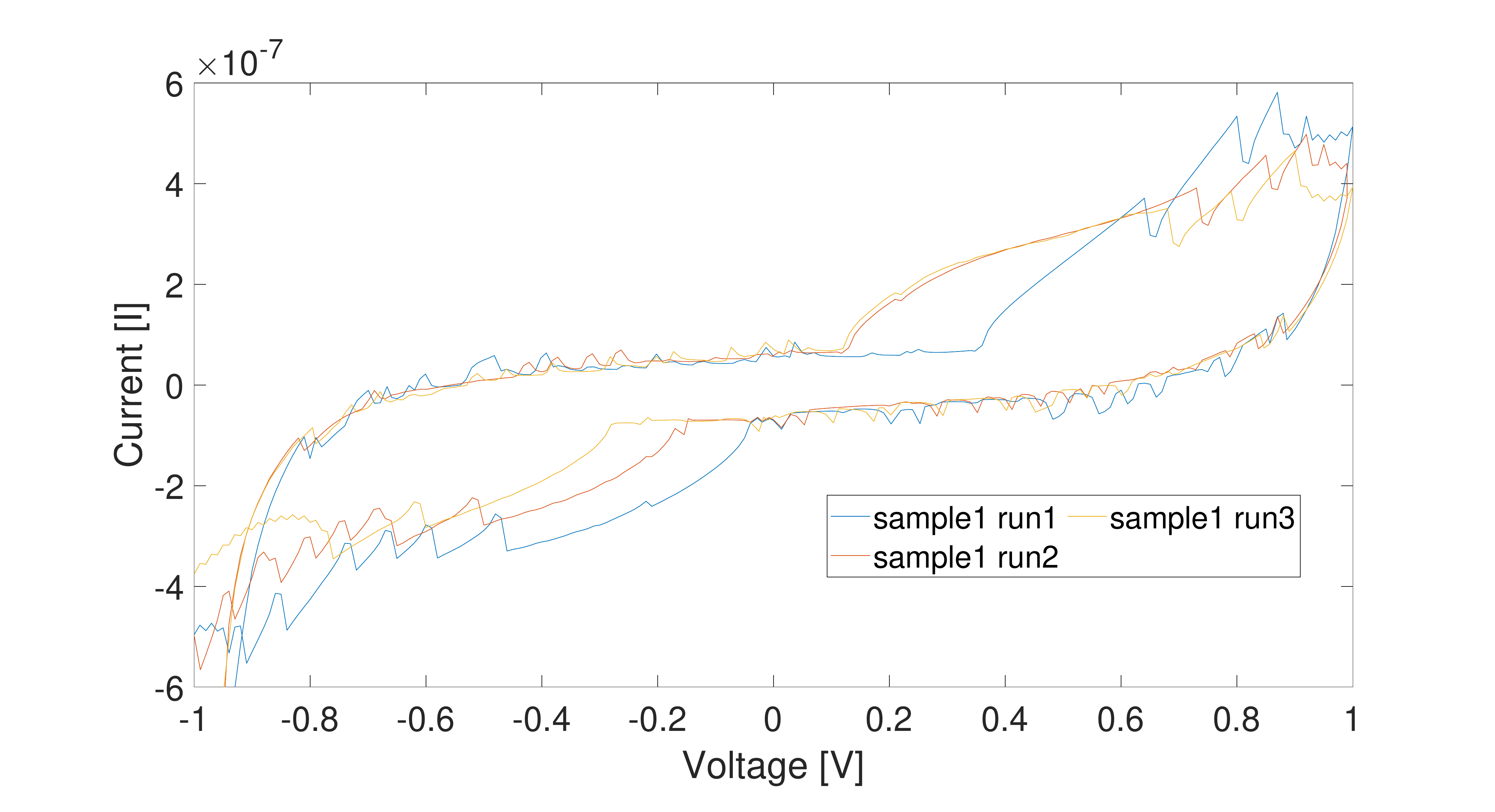}
    \caption{Cyclic voltammetry of grey oyster mushrooms. Run 1 shows the response when the sample is in the dark. Run 2 shows the response of the sample under continuous intense illumination. Runs 3 shows the response when the sample is subject to sudden changes in light conditions.}
    \label{fig:pure_fruit_bodies}
\end{figure}

Pure mycelium and fruiting bodies of the grey oyster mushroom (before functionalisation) were subject to cyclic voltammetry in intense lighting conditions, darkness and to rapid lighting changes during the cycle --- Figs.~\ref{fig:pure_mycelium} and \ref{fig:pure_fruit_bodies}. Cyclic voltammetry of the mycelium specimen under instantaneously changing light conditions (runs 3--5 in Figs.~\ref{fig:sample_1_mycelium} and \ref{fig:sample_2_mycelium}) showed no notable change when compared to  constant light conditions. Similarly, we saw the same when comparing the cyclic voltammetry of fruiting bodies in Fig.~\ref{fig:pure_fruit_bodies}. Runs 1 and 2 are from constant lighting conditions and for run 3 the light was allowed to change instantaneously. \par

\begin{figure}[!tpb]
    \centering
    \subfigure[]{\includegraphics[width=0.48\textwidth]{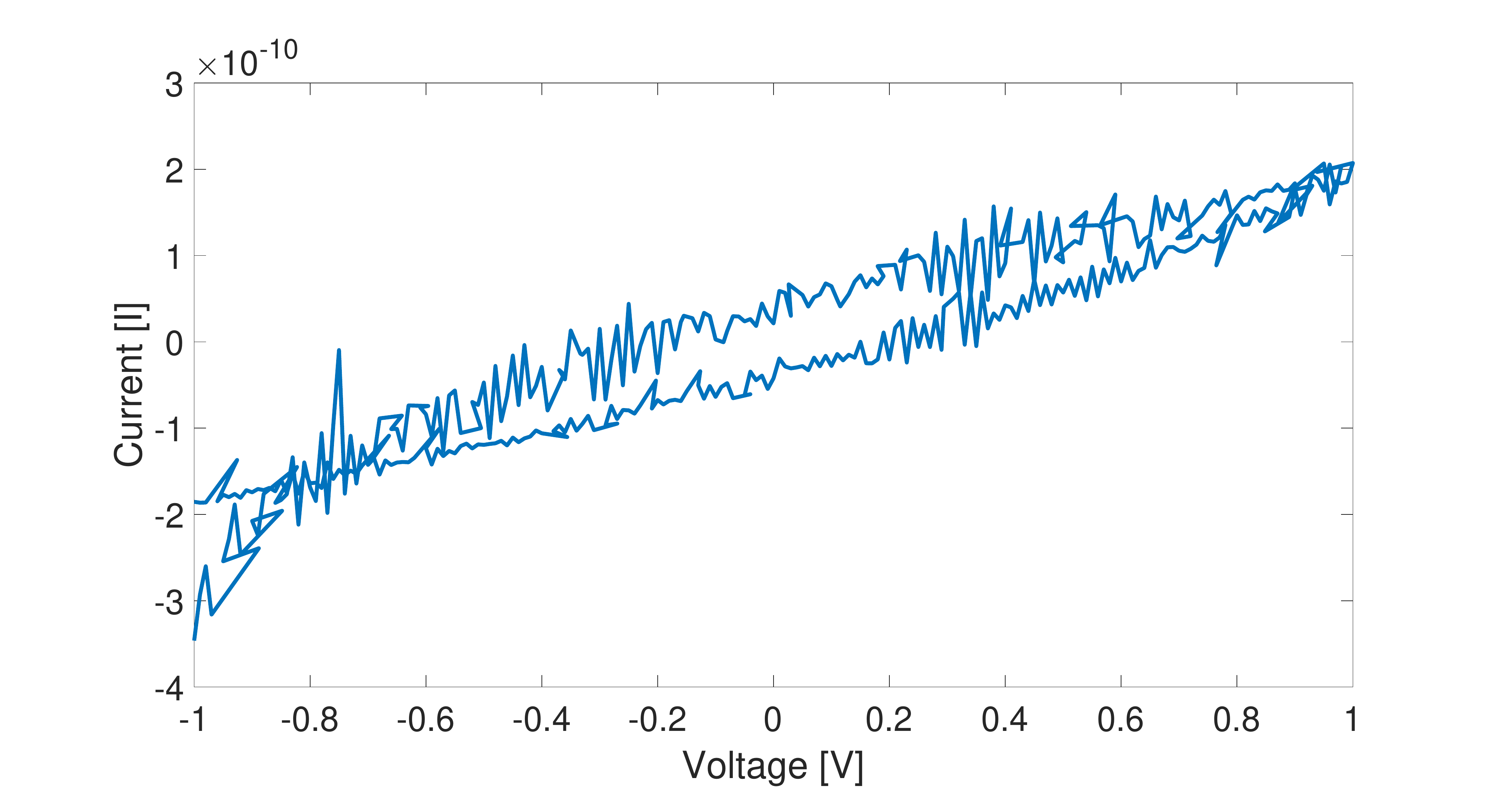}}
    \subfigure[]{\includegraphics[width=0.48\textwidth]{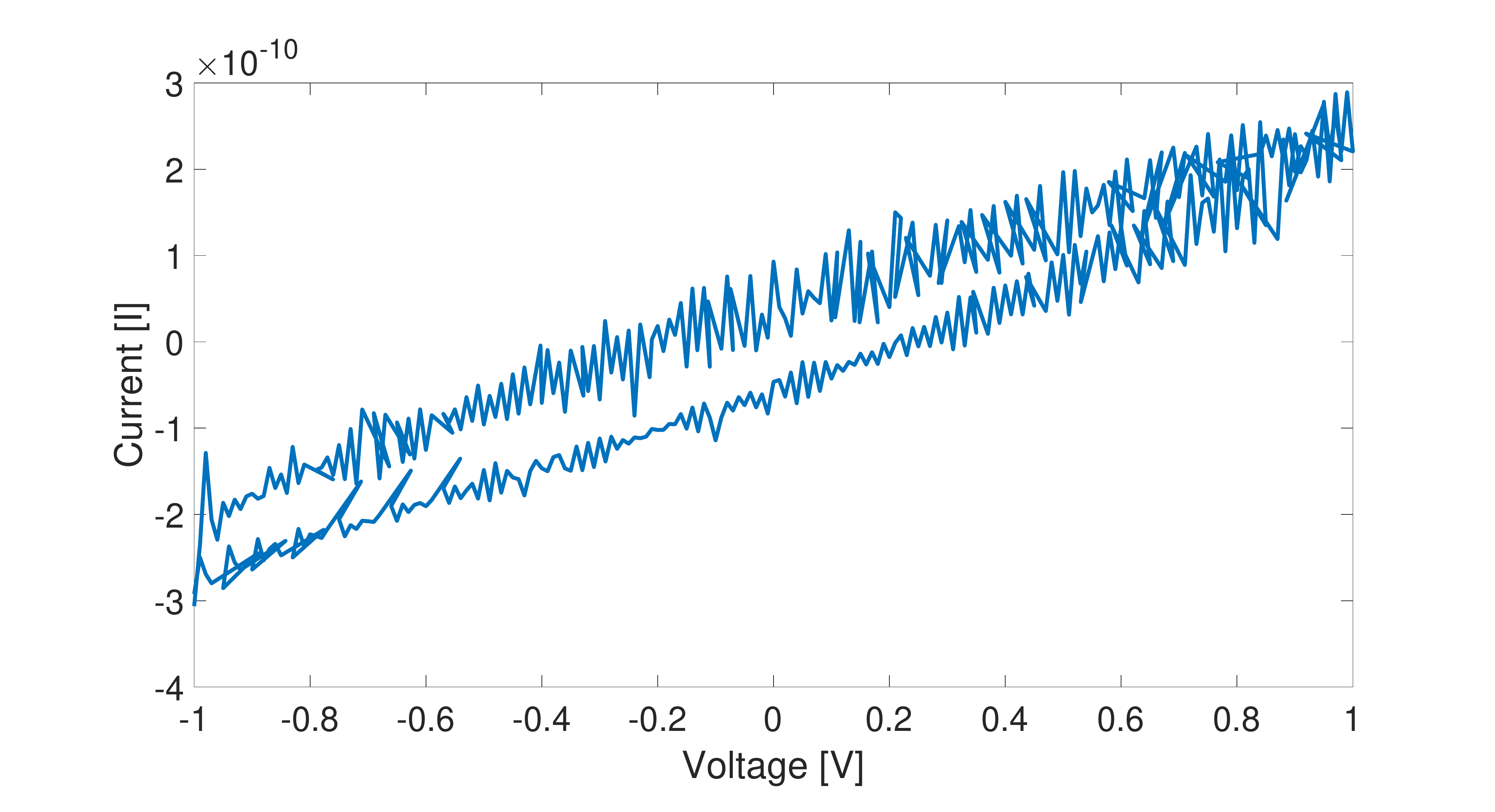}}
    \caption{Cyclic voltammetry of grey oyster mushrooms functionalised with PEDOT. (a) specimen kept in darkness. (b) specimen illuminated with 1500\,Lux light.}
    \label{fig:pdot_control}
\end{figure}

Cyclic voltammetry of fruit bodies functionalised with PEDOT:PSS demonstrates an appreciable response to sudden and dramatic changes in lighting conditions. Figure~\ref{fig:pdot_control} shows the I-V response of the functionalised fruiting body with constant light changes. The plot is a typical memfractance response that can be expected from a natural substrate~\cite{beasley2020memristive}. 

\begin{figure}[!tpb]
    \centering
    \includegraphics[width=0.48\textwidth]{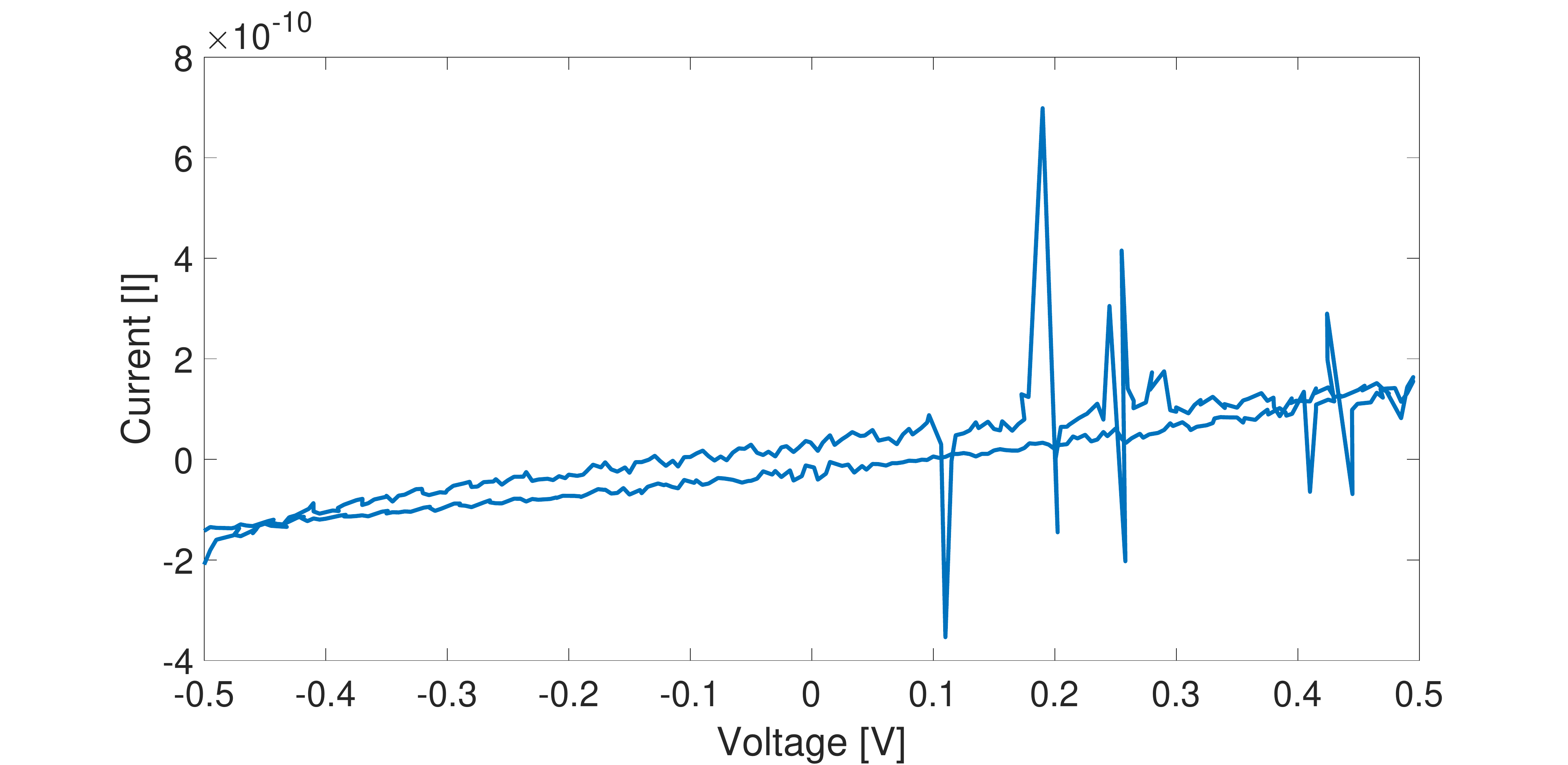}
    \caption{The fungus functionalised with PEDOT:PSS exposed to instantaneous changes between darkness and ambient lab lighting conditions.}
    \label{fig:pedot_amb_light}
\end{figure}

\begin{figure}[!tpb]
    \centering
    \subfigure[]{\includegraphics[width=0.48\textwidth]{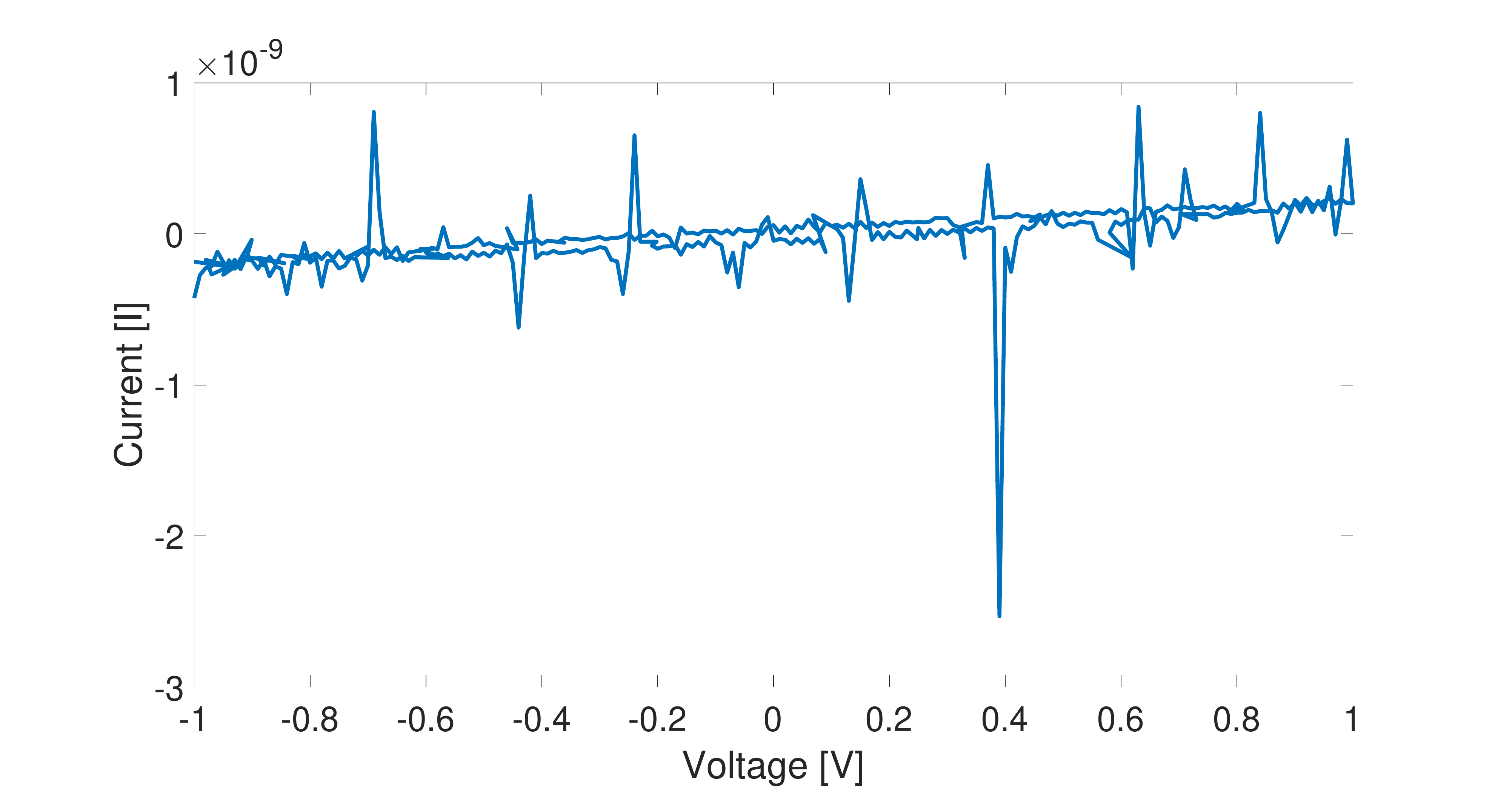}
    \label{fig:pdot_lighting_change_a}}
    \subfigure[]{\includegraphics[width=0.48\textwidth]{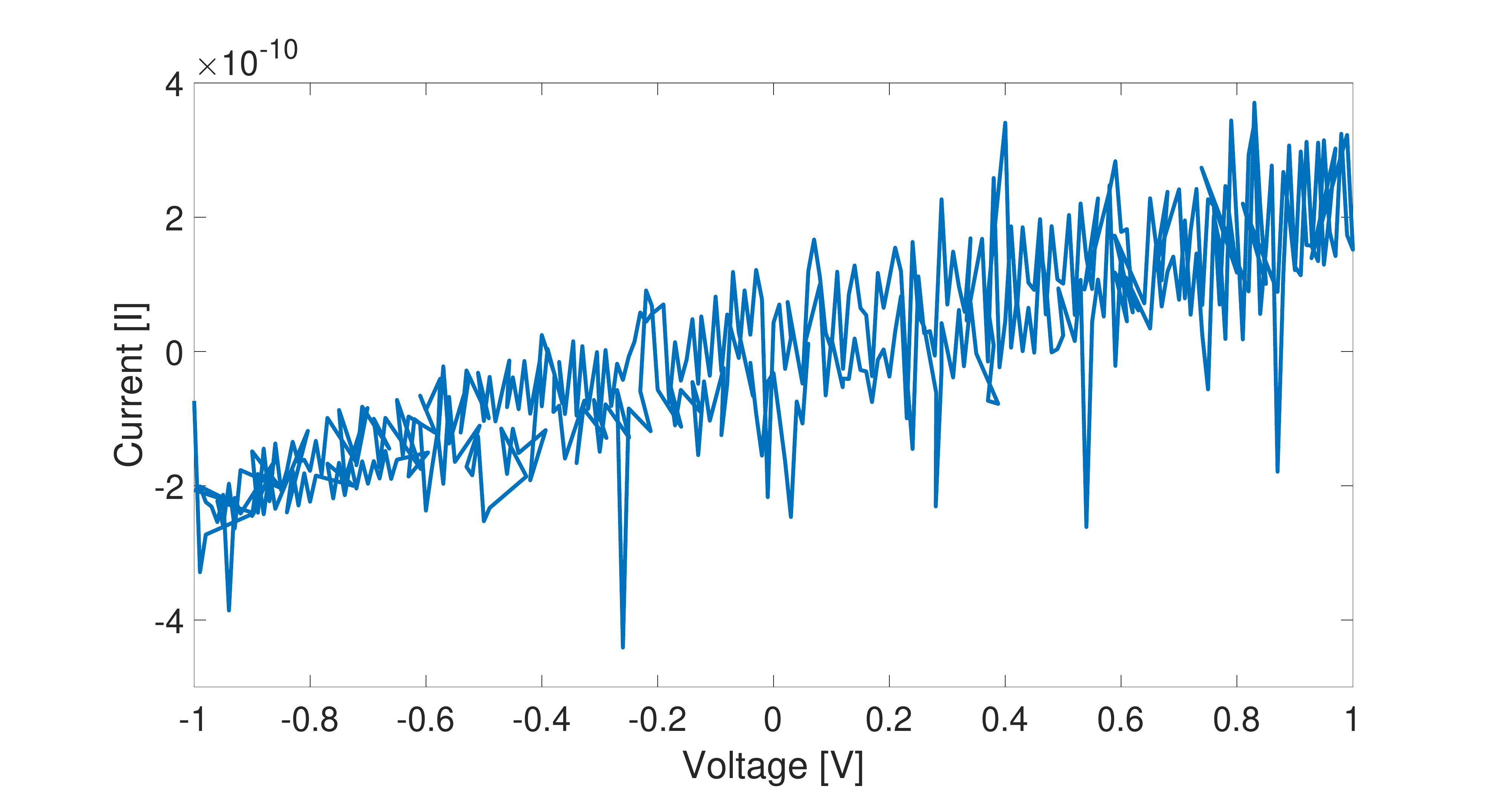}
    \label{fig:pdot_lighting_change_b}}    
    \subfigure[]{\includegraphics[width=0.48\textwidth]{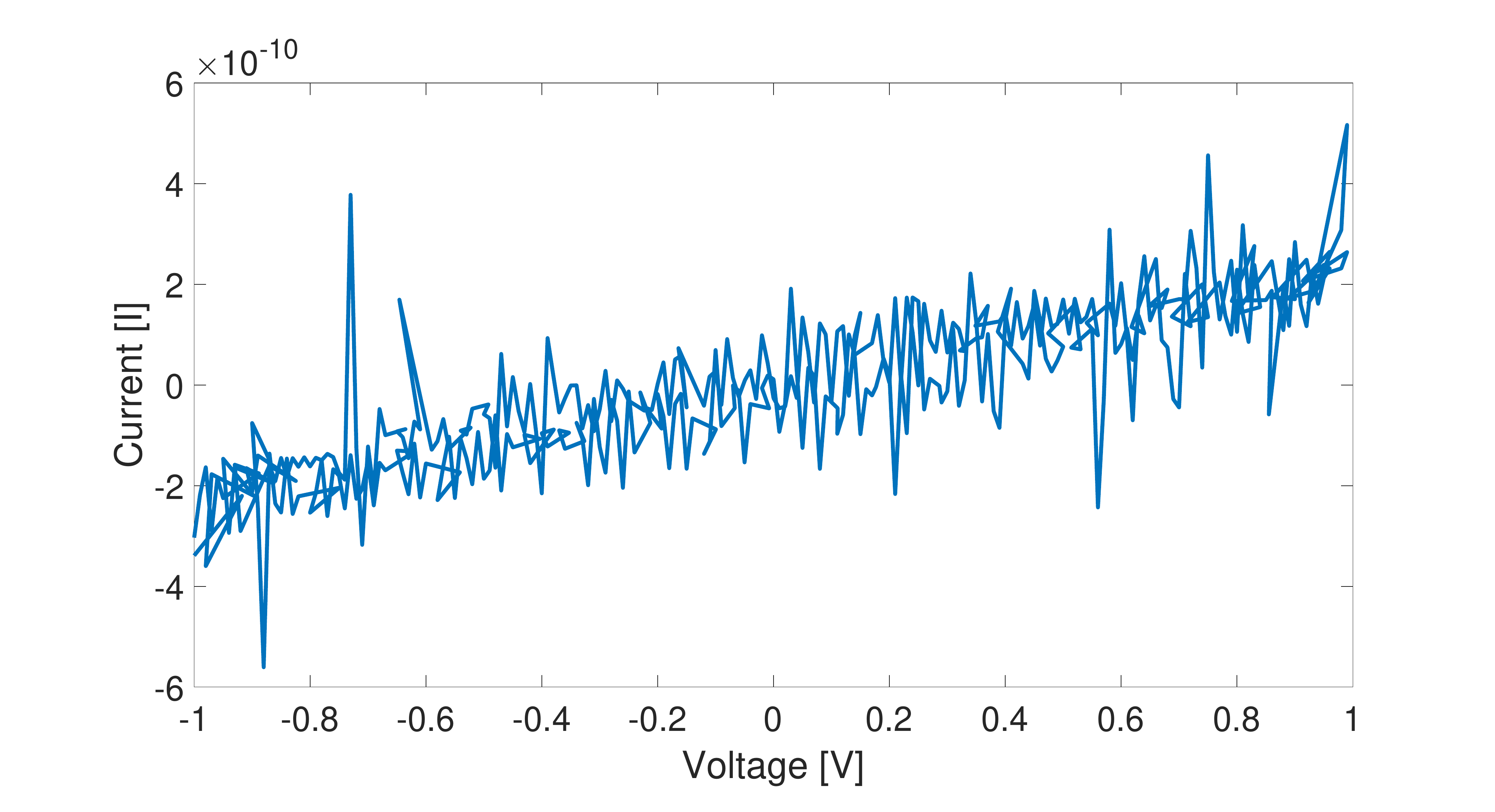}
    \label{fig:pdot_lighting_change_c}}
    \caption{Cyclic voltammetry of fungi functionalised with PEDOT:PSS. Samples are periodically exposed to intense light changes from darkness to 1500\,lux.}
    \label{fig:pdot_light_changing}
\end{figure}

In fruiting bodies injected with PEDOT:PSS, rapid changes in lighting conditions produced instantaneous changes in the conducted current. Figs.~\ref{fig:pedot_amb_light} and \ref{fig:pdot_light_changing} show the I-V response for the functionalised specimen, while the light conditions are changing. Figure~\ref{fig:pedot_amb_light} has the specimen being subject to changes in ambient lab lighting during the positive phase of the cyclic voltammetry between 0\,V and 0.5\,V. There are significant spiking results on the cyclic voltammetry curve, increasing the magnitude of the conducted current by approximately eight fold.\par

\begin{figure}[!tpb]
    \centering
    \includegraphics[width=0.3\textwidth]{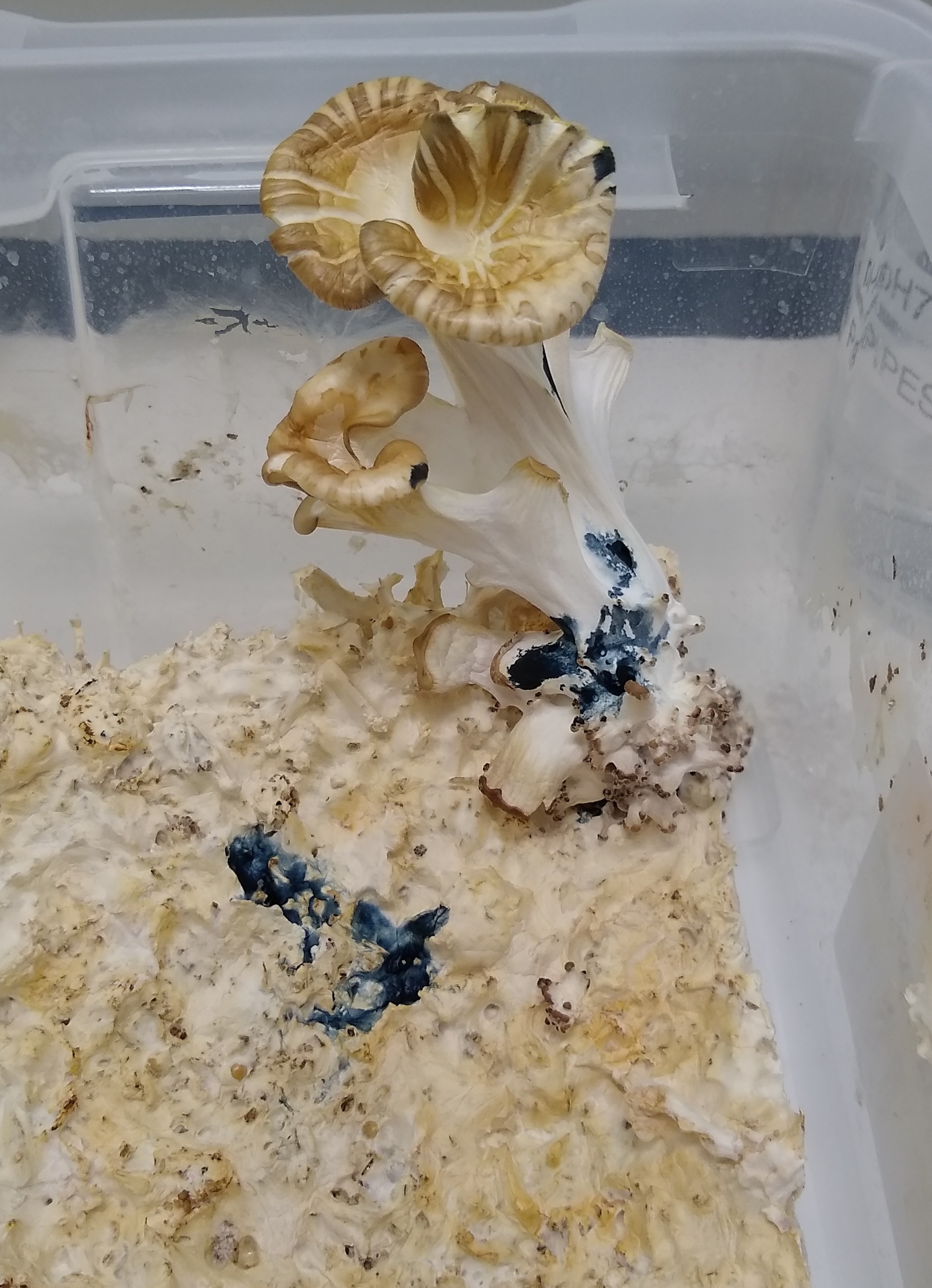}
    \caption{Dried fruit bodies and substrate colonised by mycelium. Areas of PEDOTT:PSS spillages out of stock and substrate are visible as darker areas.}
    \label{fig:dried}
\end{figure}

Figure~\ref{fig:pdot_light_changing} shows the same specimen being subjected to more extreme changes in lighting conditions. The I-V curves continue to exhibit a number of spikes, where the spikes correspond to the point at which the lighting conditions were changed. This is most clear on Fig.~\ref{fig:pdot_lighting_change_a}. Figures~\ref{fig:pdot_lighting_change_b} and \ref{fig:pdot_lighting_change_c} do still demonstrate this spiking response to the lighting stimulus, however the overall I-V curve has more noise which makes distinguishing the response harder (SNR ranges between c. 5.5 and 1.3). The injected specimen was allowed to progressively dry (Fig.~\ref{fig:dried}) out as cyclic voltammetry was conducted. The drier the specimen becomes, the noisier the recordings can become. \par

\begin{figure}[!tpb]
    \centering
    \subfigure[]{\includegraphics[width=0.48\textwidth]{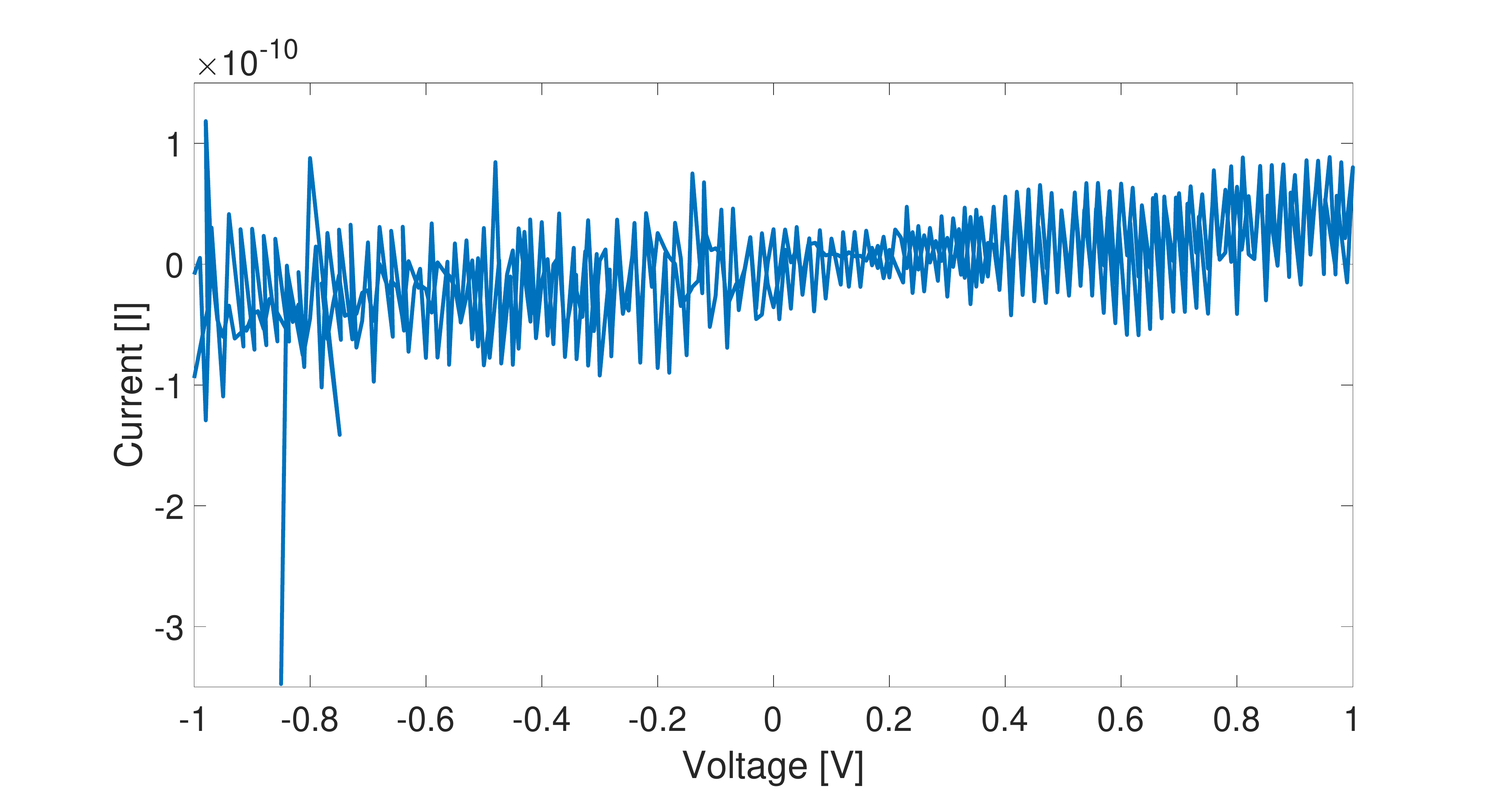}}
    \subfigure[]{\includegraphics[width=0.48\textwidth]{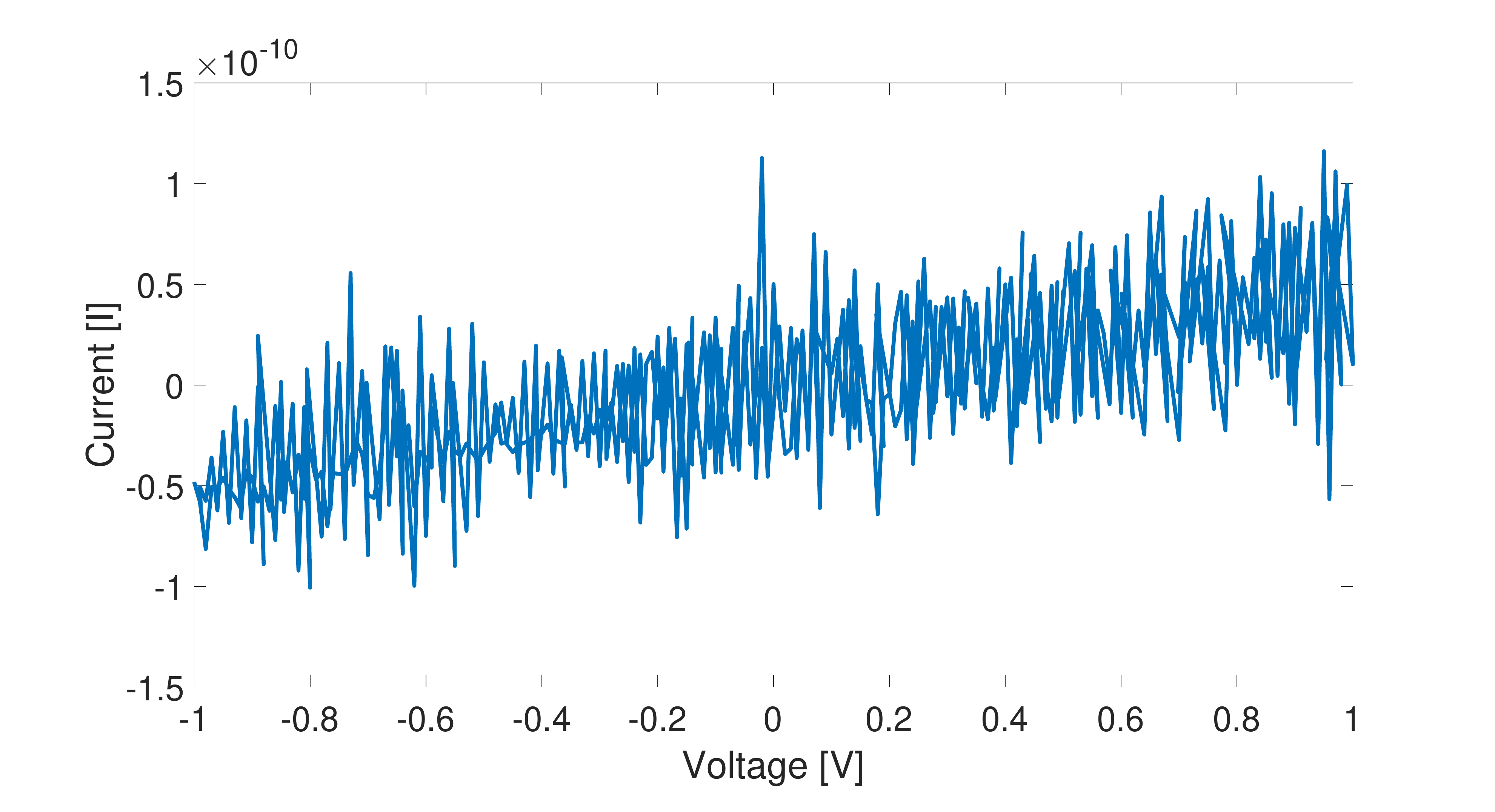}}
    \caption{Cyclic voltammetry of dried sample of grey oyster mushrooms. (a) Specimen kept in darkness. (b) Specimen illuminated with 1500\,Lux light.}
    \label{fig:dry_control}
\end{figure}

Dried fruit bodies located further, c.~25\,cm, from the injection site of the PDOT:PSS were also subjected to cyclic voltammetry. Figure~\ref{fig:dry_control} shows the response of the sample under the control conditions --- sample kept in darkness and sample exposed to light. The specimen had very poor conductive properties. Presumably this is due to the fact that the injection site of the PEDOT:PSS was sufficiently far from the recording site that there was, at best, only minimal uptake of the conductive polymer. However, sweeping the voltage did result in some change in conducted current, although always in the pico-Amps range.

\begin{figure}[!tpb]
    \centering
    \subfigure[]{\includegraphics[width=0.48\textwidth]{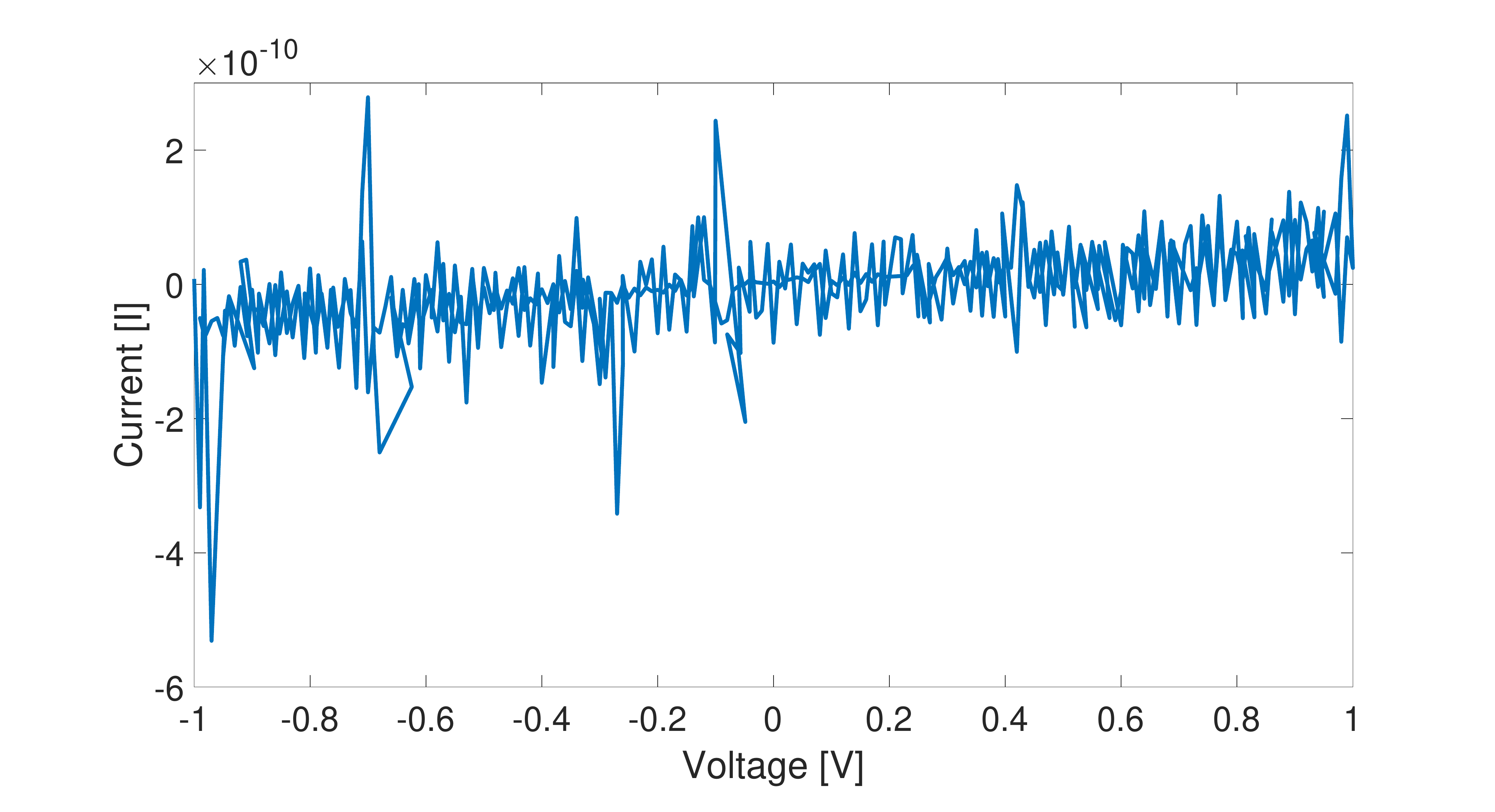}}
    \subfigure[]{\includegraphics[width=0.48\textwidth]{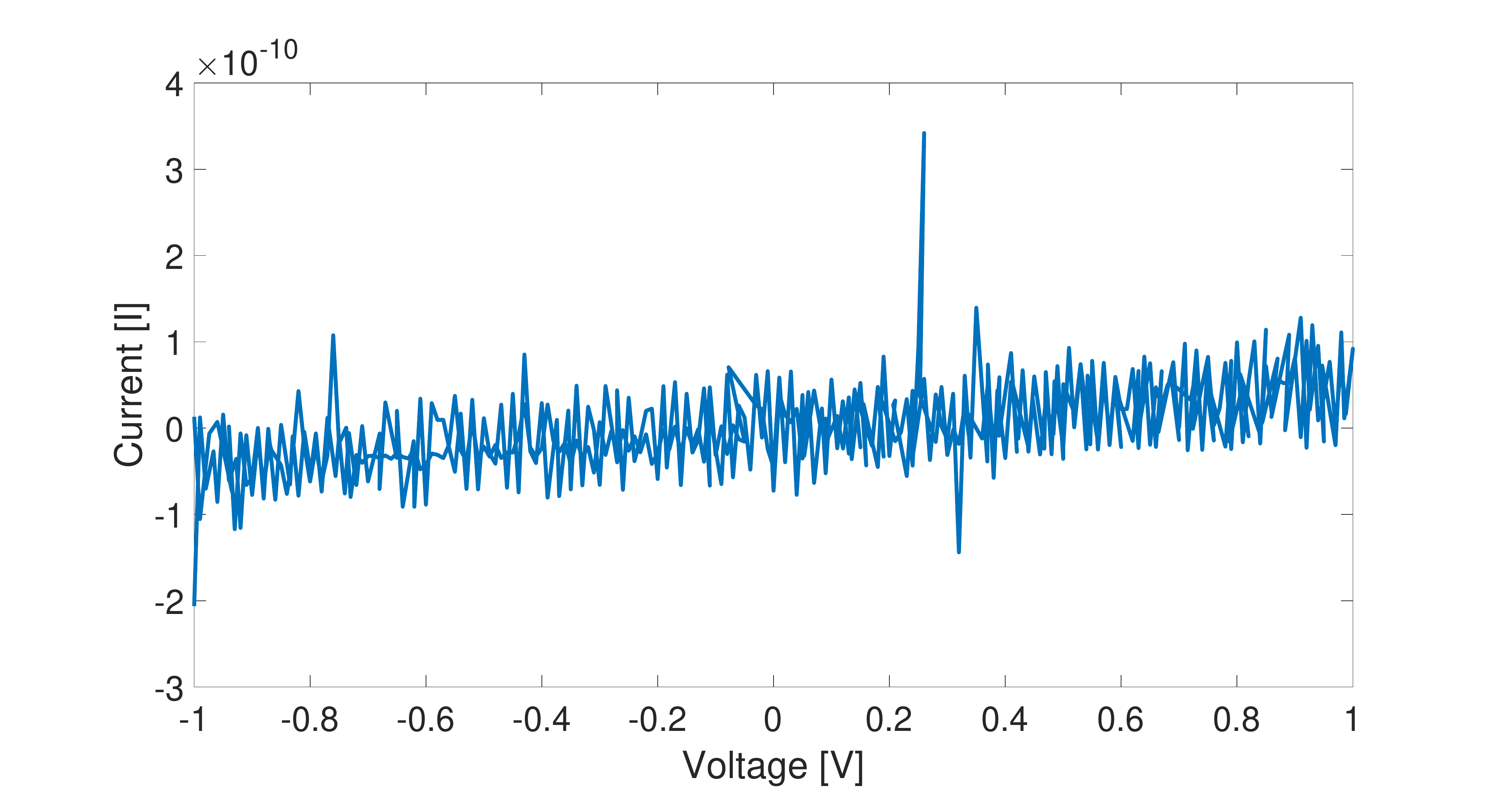}}
    \caption{Cyclic voltammetry of dried sample of grey oyster mushrooms. Samples are periodically exposed to intense light changes from darkness to 1500\,lux.}
    \label{fig:dry_lighting_change}
\end{figure}

Cyclic voltammetry was then performed on the same specimen during intense changes in lighting conditions (Fig.~\ref{fig:dry_lighting_change}). Here, increasing and decreasing the light levels also resulted in an instantaneous spike in the conducted current. However, this specimen was not as responsive as the specimen injected directly with PEDOT:PSS.

\section{Conclusions}
\label{sec:conclusions}

Fungi are  known for their photosensitive nature, with research showing fungi can be more photosensitive than green plants~\cite{Carlile1965fungiphotobiology, Furuya1986fungiphotobiology}. Indeed, fungi rely on different lighting conditions to control their reproduction cycle. Fungi are most receptive towards the UV end of the spectrum but exhibit photoresponses across the entire light spectrum. Briefly exposing fungi to light can interrupt their current growth cycle, triggering other responses. The reported photosensitivity of fungi inspired us to check if the fungi can be used as photosensors.

The cyclic voltammetry of the mycelium and fruit bodies specimens shows that exposing the test subject to instantaneously changing light conditions has no discernible impact on conducted current when compared to keeping the light conditions constant. Extreme constant lighting conditions (darkness or extreme illumination) again has little to no bearing on the I-V curves for cyclic voltammetry. Under all conditions, the specimens exhibit characteristic mem-fractance responses\cite{beasley2020memristive}. The lack of response to the stimulation with light might be due to the fact that the biochemical responses required to initiate electric responses might take several hours and, therefore, the length of recordings made here was not sufficient to capture such changes. This leads to us to conclude that intact fungi can be not be used as immediate-response photoswitches. 

Functionalising the specimen with a conductive polymer,  in this case PEDOT:PSS, and re-performing cyclic voltammetry yielded a photosensitive response when the light conditions change instantaneously. Therefore, in line with previous research~\cite{lin2013self}, polyfluorene and PEDOT:PSS photosenesing inks~\cite{lavery2011all}, n-Si/PEDOT:PSS solar cells~\cite{pietsch2014interface}, by functionalising mushrooms using the conductive polymer PEDOT:PSS, we were able to produce a photosensor. Furthermore, the spikes in the cyclic voltammetry were significant enough to be detectable by additional hardware. Future work would be to develop the digital back-end that can be used to interpret and process the spiking response of the mushrooms.

\section*{Acknowledgements}
This project has received funding from the European Union's Horizon 2020 research and innovation programme FET OPEN ``Challenging current thinking'' under grant agreement No 858132.

\section*{Author contributions}
A.B. conceived the idea of experiments. A.A. and A.P. prepared the substrate colonised by mycelium. A.B. performed experiments, collected data and produced all plots in the manuscript.
A.B. and A.A. prepared manuscript (wrote and reviewed all contents).
A.P. reviewed manuscript.

\bibliographystyle{plain}
\bibliography{references_capacitors, references, references_photosensor}

\end{document}